\documentclass{comnet}
\usepackage{ragged2e}
\usepackage{subcaption}
\usepackage{array} 
\usepackage{graphicx}
\setcitestyle{numbers}
\usepackage{enumitem}
\usepackage{float}
\usepackage{amssymb}
\usepackage{amsmath}
\usepackage{hyperref}

\begin{document}

\title{Analyzing Communicability and Connectivity in the Indian
Stock Market During Crises}

\shorttitle{Communicability and Connectivity in the Indian
Stock Networks} 
\shortauthorlist{Pawanesh Pawanesh, Charu Sharma, Niteesh Sahni} 

\author{
\name{Pawanesh Pawanesh$^*$}
\address{Department of Mathematics, Shiv Nadar Institution of Eminence Deemed to be University, Dadri, Gautam Buddha Nagar, Uttar Pradesh, India \email{$^*$Corresponding author: py506@snu.edu.in}}
\name{Charu Sharma}
\address{Department of Mathematics, Shiv Nadar Institution of Eminence Deemed to be University, Dadri, Gautam Buddha Nagar, Uttar Pradesh, India}
\and
\name{Niteesh Sahni}
\address{Department of Mathematics, Shiv Nadar Institution of Eminence Deemed to be University, Dadri, Gautam Buddha Nagar, Uttar Pradesh, India}}

\maketitle

\begin{abstract}
{Understanding how information flows through the financial networks is important, especially during times of market turbulence. Unlike traditional assumptions where information travels along the shortest paths, real-world diffusion processes often follow multiple routes. To capture this complexity, we apply communicability, a network measure that quantifies the ease of information flow between nodes, even beyond the shortest path. In this study, we aim to examine how communicability responds to structural disruptions in financial networks during periods of high volatility. We compute communicability-based metrics on correlation-derived networks constructed from financial market data, and apply statistical testing through permutation methods to identify significant shifts in network structure. Our results show that approximately 70\% and 80\% of stock pairs exhibit statistically significant changes in communicability during the global financial crisis and the unprecedented COVID-19 crisis, respectively, at a significance level of 0.001. The observed shifts in shortest communicability path lengths offer directional cues about the nature and depth of each crisis. Furthermore, when used as features in machine learning classification models, communicability measures outperform the shortest-path-based measures in distinguishing between market stability and volatility periods. The performance of geometric measures was also comparable to that of topology-based measures. These findings offer valuable insights into the dynamic behavior of financial markets during times of crises and underscore the practical relevance of communicability in modeling systemic risk and information diffusion in complex networks.}
{Financial network analysis, Complex network, Communicability measure, Support vector machine}
\\
\end{abstract}

\section{Introduction}
\justifying
The financial system is widely regarded as one of the most complex systems. At its most fundamental structural level, the global financial network comprises billions of individual transactions and organizations, including traders, institutions, and markets, communicating via financial instruments and market interactions. These entities are linked within localized institutions, such as regional financial markets or sectors, and then aggregated into bigger systems, such as the global stock exchange. Researchers are studying this system using methods and tools from across fields to help better understand and advance economic growth and development, policy-making and regulation, risk management and sustainability, technological advancement, and innovation. In the recent past, modelling a financial system as an interconnected network of nodes has been a popular approach. In particular, complex networks have emerged as a powerful tool for better modelling the pairwise interaction between financial entities such as stocks, cryptocurrency, and bonds. Many researchers have modelled the pairwise correlations between daily logarithmic returns of stocks to construct the networks. Kendall’s tau and mutual information measures have also been widely used to model the interaction between stock pairs. A substantial amount of literature leverages the topology and geometry of these financial networks \cite{ansari2024novel, guo2018development,jiang2014structure, keller2021hyperbolic, macmahon2015community,nie2021studying,pawanesh2025exploring, pawanesh2025exploiting,purqon2021community, samitas2020machine}. In a similar line, several authors have used traditional network metrics such as Eigenvector centrality, betweenness centrality, degree centrality, and hybrid centrality measures to study these financial networks to propose portfolio selection strategies, risk management, and uncover hidden patterns in stock market dynamics \cite{ansari2024novel, borgatti2005centrality,jiang2014structure,pozzi2013spread}.

In the past literature, it has been observed that the study, prediction, and identification of market stability and volatility periods have been crucial research topics in financial Markets. Many authors have studied and focused on the correlation matrix of stock returns and captured the market turbulence periods \cite{araujo2007geometry, echaust2014geometry,mendes2003reconstructing,vilela2002process}. Furthermore, several authors have used network theory tools on the correlation matrix to detect and predict the stock market crisis periods \cite{kukreti2020perspective, kulkarni2024investigation, samitas2020machine, xiu2021crash}. For instance, in 2020, Kukreti et al. \cite{kukreti2020perspective} studied the correlation-based networks and used recently developed entropy measures, such as structural entropy and eigen entropy, and showed how entropy measures could be used to identify normal, bubble, and crash periods. A popular literature \cite{pharasi2018identifying,pharasi2019complex} also explores the random matrix theory to track the market states and long-term precursors of crises. It proposes that during times of crisis, the whole financial market behaves in a way that acts like a single huge community and cluster. In later years, Chakraborti et al. \cite{chakraborti2020phase} have revealed the distinct market phases such as crisis, bubble, and anomalies using the combined framework of the eigen decomposition and eigen vector centrality. They highlight that the phase separation and order-disorder transition are robust tools to monitor the systemic risk across the distinct periods of the stock market. Kulkarni et al. \cite{kulkarni2024investigation} also studied the correlation-based network of stocks comprising the National Stock Exchange (NSE) and the Bombay Stock Exchange (BSE) using geometry-inspired measures of network. They captured the crisis periods of the Indian stock market. Specifically, they monitored the changes occurring in the two Indian stock markets by analyzing the fluctuations in standard network measures, discrete Ricci curvatures, and persistent homology-based topological measures. As we mentioned earlier, several network metrics are well-documented in the literature, and many of them were used to study financial networks and propose a deeper understanding of the system.

Thus, according to the literature discussed above, network metrics have been crucial tools for investigating these challenges over the full cross-correlation matrix and the correlation-based network. When working with the network’s shortest path-based approaches, the authors primarily explored traditional network measures, yet limited conclusive results were reached. It would be exciting to investigate the potential of alternative, unconventional network metrics as novel and effective in distinguishing market stability and volatility periods. We know that stock networks are more connected during times of crisis, indicating larger linkages between equities as market dynamics change. This is frequently caused by panic selling, elevated volatility, and the spread of systemic risk.

In the recent literature, we have observed the application of the communicability measure of the networks in biology and world trade networks \cite{lella2019communicability,lella2020communicability}. Thus, communicability is a broader measure of connectedness that seeks to quantify the ease of communication between two nodes by considering not only the shortest path but all conceivable paths linking them \cite{crofts2009weighted, estrada2008communicability}. This measure indicates the network property of information to flow under a diffusion model: as a result, it could be particularly suited to studying the turbulent periods networks and may be highly sensitive. Lella et al. \cite{lella2019communicability} have explored and used the communicability measure to study the connectivity disruption among brain regions caused by white matter degeneration due to Alzheimer’s disease. Later, Lella and Estrada \cite{lella2020communicability} used the communicability distance to reveal the hidden pattern in patients with Alzheimer’s disease. First, they show that the shortest communicability path length performs noticeably better than the shortest path length in distinguishing between people with Alzheimer’s disease (AD) and healthy persons. Additionally, they identified structural elements that appear to be responsible for the onset of AD, specifically, the areas of the brain where AD significantly impacts communicability distance. Most of them connect the vermis or both hemispheres of the brain. In the same year, Bartesaghi et al. \cite{bartesaghi2020community} studied the mesoscale structure of the World Trade Network, identifying clusters using vibrational and communicability distances. The approach is more computationally efficient than traditional clustering techniques, identifies unique country links, and reveals inter- and intra-cluster features. Their recent work (Bartesaghi et al.) \cite{bartesaghi2020risk, bartesaghi2025global} derived the communicability measure from the matrix exponential and effectively applied it in the financial network analysis. They proposed a risk-sensitive centrality measure based on communicability to identify key economic and financial agents. More recently, in \cite{bartesaghi2025global}, they explored global balance and systemic risk through communicability in correlation-based financial networks. These studies affirm the suitability of communicability-based tools for uncovering hidden vulnerabilities and systemic connections within the economic systems.

On the other hand, a recent paper \cite{keller2021hyperbolic} demonstrates that the hyperbolic space is suitable for studying and understanding the underlying dynamics of these financial networks. Specifically, they showed that the hyperbolic geometry can well represent the European banking network in hyperbolic space, a geometry of negative curvature. Authors propose that banks' positions reflect their systemic importance and regional affiliations, captured through ``popularity'' and ``similarity'' dimensions in the hyperbolic space. Interestingly, while core institutions remained stable over time, the peripheral structure evolved, revealing how effectively hyperbolic embedding of the complex networks can help track more profound structural changes in financial systems. The previous studies of the complex network following the heterogeneous degree distribution in the hyperbolic space also support the idea that the underlying geometry of the network boosts the capability of the network-based algorithms to capture the hidden structure of the system \cite{cacciola2017coalescent, keller2021hyperbolic, longhena2024detecting,longhena2025hyperbolic, muscoloni2017machine, zhao2023spatial}. A particularly effective hyperbolic embedding algorithm, the coalescent embedding technique \cite{muscoloni2017machine}, has been very popular for embedding complex networks in 2D or 3D hyperbolic spaces. This algorithm uses the network power-law property to embed the complex network in the hyperbolic space efficiently. Initially, this algorithm has been applied in biological and neural systems \cite{cacciola2017coalescent, longhena2024detecting, longhena2025hyperbolic}, where authors detect subtle shifts in brain connectivity, revealing region-specific disruptions linked to conditions such as epilepsy and Alzheimer’s. These studies demonstrate that this geometric approach can surpass traditional graph-theoretic tools in identifying affected areas with greater clarity and diagnostic relevance. Inspired by these breakthroughs, our recent work (Pawanesh et al. 2025) \cite{pawanesh2025exploiting} extended the use of coalescent embedding to the Indian stock market, modelling it as a scale-free network within hyperbolic space. The findings reveal that the underlying geometric structure enhances community detection, accurately identifies market sectors, and provides early warning signals of volatility. This motivates our use of coalescent embedding to capture the geometric structure of the financial network and exploit it for a classification framework.

In the present manuscript, we used a dataset of the NIFTY 500 Index from the National Stock Exchange (NSE), India, corresponding to different financial events. The first step of our analysis is to model the stock market data as a complex network by considering pairwise correlations between stocks. It is worth noting that the full correlation-based network has a homogeneous degree distribution. In contrast, filtered networks such as Minimum Spanning Tree (MST) and the Planar Maximally Filtered Graph (PMFG) subnetworks preserve the most statistically significant correlation structure, while mitigating the redundant connections from the full network. Thereby, offering us a more meaningful and optimal structure for analysis. The MST and PMFG networks have been popular models to explore in a huge body of research over the years \cite{ansari2024novel, guo2018development,jiang2014structure, macmahon2015community,nie2021studying,nie2018constructing,sharma2019mutual,tumminello2005tool}. In our previous work \cite{pawanesh2025exploiting}, we have already established that these subgraphs exhibit the power-law degree distribution. Furthermore, in this manuscript, we will restrict our analysis to the PMFG network because this maintains several pathways, maintains structural complexity, and facilitates communication to capture indirect interactions, network resiliency, and realistic dynamics, providing deeper insights into network behaviour.

Next, we begin our analysis to determine how well the communicability network distinguishes the market stability and volatility periods. Thus, we used the shortest path-based measures, namely edge betweenness centrality, shortest path length, and communicability measure, as Features in the support vector machine (SVM) \cite{james2013introduction} to investigate how well it classifies the market periods. The results show that, statistically, communicability is more sensitive to periods of volatility than the shortest path length of the financial network. Additionally, these network metrics, edge betweenness centrality (EBC), shortest path length (SPL), and communicability (COMM), individually are able to classify the market periods of volatility using features in the SVM classifier with 90\% accuracy. Furthermore, we used the SVM classifier on the geometry-based metrics of the embedded network in the hyperbolic space. The results suggest that those geometric measures perform comparably to network topology-based measures.

The rest of the manuscript is divided into five sections. In section 2, we describe the data we utilized in our analysis. Next, in section 3, we briefly overview the preliminary definitions and methodology. In section 4, we give a detailed explanation of the results. Section 5 discusses and highlights the silent observations and limitations. Lastly, in the concluding section, we highlight our main findings and the work’s implications for the future.

\section{Data description}
\justifying

We analyze daily closing price data from two major stock market crises: the standard global financial crisis of 2008 and the unprecedented pandemic-induced crisis of 2020. For the 2008 crisis, we focus on all 500 stocks listed in the NIFTY500 index (a benchmark index representing the top 500 companies listed on the National Stock Exchange (NSE), India), using data from the years 2005 and 2008. These two periods represent contrasting market conditions, with 2005 characterized by relatively low volatility and 2008 by heightened volatility. Some of the stocks were removed from the analysis because of the missing price data. This leaves us with the complete data of 218 stocks in both years. Next, we calculate the daily logarithmic returns \(r_{i}(t) = \log (p_{i}(t+1)) - \log (p_{i}(t))\). Here, \( p_{i}(t) \) stands for the closing price of the stock \( i^{th} \) on day \( t \). During this, we remove the data corresponding to non-successive days(like Friday \& Monday are separated by more than 1 day due to holidays in between). Thus, we ignored the log return for Monday; similarly, there were other exceptions because of bank holidays. We are left with 191 logarithmic return values for the year 2005 and 183 logarithmic return values for the year 2008, corresponding to each of the selected 218 stocks in both years. 

Following a comparable methodology and aligning the temporal gap, we selected 2017 and 2020 to represent periods of market stability and volatility, respectively, in the context of the 2020 pandemic crisis. These years were characterized by relatively low (2017) and high (2020) standard deviations in market returns. After data pre-processing and filtering, we retained a consistent 383 stocks for both years. This yielded 193 logarithmic return observations for 2017 and 196 for 2020, corresponding to each 383 stocks in their respective time frames.

The above selection of specific years in this manuscript was made intentionally to maximize the contrast between distinct phases of the financial market. Rather than selecting consecutive years, we chose adjacent periods that represent stable (2005) and volatile (2008) market conditions, thereby enabling a meaningful comparison of structural differences in financial networks during contrasting financial events. Similarly, the pair of years (2017, 2020) was selected to analyze the structural impact of a second unprecedented market crisis.
 
\section{Methods and preliminaries}
This section first discusses the correlation-based network formulation using the above datasets. Then, we briefly overview traditional complex network measures such as edge betweenness centrality (EBC), shortest path length (SPL), average clustering coefficient (ACC), and communicability measure (COMM). In addition, we provide an overview of the algorithmic procedure of the permutation test, which we used to capture the significant stock pairs in the financial markets during significant market events.

First, we introduce a dependency measure known as the Pearson correlation coefficient (PCC), which quantifies the linear dependence between two stocks $i$ and $j$ is denoted as $C=[C_{ij}]_{N\times N}$, where $N$ is the number of stocks. Mathematically, PCC is defined as follows:
\begin{equation}
    C_{ij} =  \frac{E[r_{i}r_{j}]-E[r_{i}]E[r_{j}]}{\sqrt{E[r_{i}^2]-E[r_{i}]^2}\sqrt{E[r_{j}^2]-E[r_{j}]^2}} 
    \label{eq:1}
\end{equation}
Where $r_i$ and $r_j$ are the log returns of the stock $i$ and stock $j$, respectively, and $E(r_{i})$ represents the expected value of returns corresponding to the $i^{th}$ stock. Additionally, to model the pairwise similarity between the stocks as a distance, we define the correlation distance as follows:
\begin{equation}
    D = [d_{ij}] = [\sqrt{2*(1-C_{ij})}] 
    \label{eq:2}
\end{equation}
The importance of using $d_{ij}$ as a weight has been highlighted \cite{aste2010correlation, guo2018development,jiang2014structure,kumar2012correlation,nie2021studying,nie2018constructing, pozzi2013spread, sharma2019mutual,tumminello2005tool}. This distance is, in fact, equivalent to the Euclidean distance between the normalized return series \cite{mantegna1999introduction}.

In this paper, we worked with the weighted and unweighted Planar maximally filtered graph (PMFG) \cite{tumminello2005tool}. We chose the weighted network representation $G = (V, E, W)$, where $V$ is the set of vertices (stock), $E$ is the set of edges, and $W$ is the weighted adjacency matrix whose element $w_{ij}$ is associated with each edge in the network. Thus, their weighted and unweighted adjacency matrices can be represented as

\begin{equation}
W = [w_{ij}] \quad \text{where} \quad
w_{ij} = 
\begin{cases} 
C_{ij}, & \text{if } i \text{ and } j \text{ are adjacent} \\ 
0, & \text{otherwise}
\end{cases}
\label{eq:3}
\end{equation}

\begin{equation}
A = [a_{ij}] \quad \text{where} \quad
a_{ij} = 
\begin{cases} 
1, & \text{if } i \text{ and } j \text{ are adjacent} \\ 
0, & \text{otherwise}
\end{cases}
\label{eq:4}
\end{equation}
respectively. We construct the PMFG network using a window of 60 days with one day as a rolling window on each year’s dataset separately for both the crisis periods. Thus, we have a total of 132, 124, 134, and 137 PMFG network windows for 2005, 2008, 2017, and 2020, respectively. Further, we provide some of the network’s basic topology metrics, such as average degree, network diameter, and clustering coefficient (Eq. ~\ref {eq:6}), highlighting the structural change and collectively capturing the network's density, connectivity, and compactness. For that, first, we have provided a network visualisation generated over a 60-day window (a randomly chosen window number 51) for the years 2005 and 2008, respectively, below in Figures ~\ref{fig1} (a) and ~\ref{fig1} (b). We plot these networks using the Gephi 0.10.1 (https://gephi.org/) tool. In Figure ~\ref{fig1} (a), the network has an average weighted degree of 4.10 and a weighted diameter of 8.65, and in Figure ~\ref{fig1} (b), an average weighted degree of 4.55 and a weighted diameter of 6.81. Similarly, we have also generated the figures for the pandemic crisis of the year 2020, provided in Appendix B (see Fig. B1). These observations reflect notable changes in network topology during periods of financial stress, suggesting heightened market interconnectedness and potential systemic risk during crisis episodes.

Next, we have also dynamically calculated the weighted network diameter and clustering coefficient for all the stable and volatile windows of that crisis event. For the relevant histogram plots of the diameters of the network windows of stable (2005), volatile (2008) periods of the global crisis and pandemic crisis, respectively, see Appendix B (Fig. B2). Similarly, Appendix B (see Fig. B3) provides the clustering coefficients of all the network windows corresponding to the stable (2005) and volatile (2008) periods of the global crisis and pandemic crisis, respectively. The figure indicates that the network diameter median decreased during the global financial crisis (2008). This reduction suggests that the market became more tightly interconnected during the height of the crisis, likely due to increased correlations among stock returns, which is a common feature during financial contagion. On the other hand, the network diameter increased during the pandemic financial crisis of 2020. This expansion indicates a more dispersed or fragmented market structure, possibly reflecting heterogeneous market reactions across sectors and geographies. The increased diameter could suggest that the market did not move as a single unified system but rather in clusters or communities with varying dynamics, possibly due to differing impacts of the pandemic on industries. The clustering coefficient histogram also supports the same interpretation provided in Appendix B.
\begin{figure}[H]
    \centering
    \begin{subfigure}{0.7\textwidth}
        \centering
        \includegraphics[width=\textwidth]{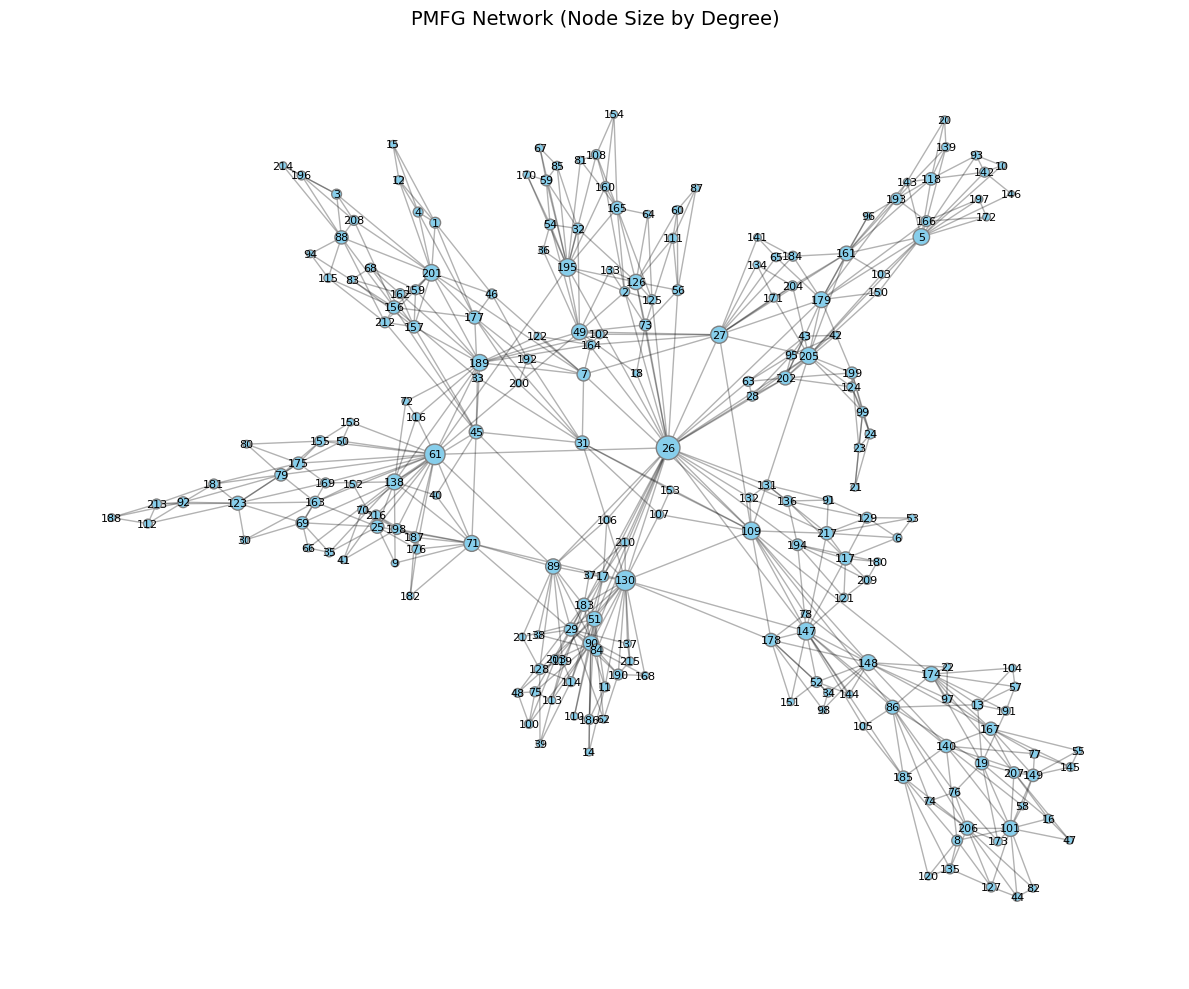}
        \textbf{(a)} 
        \label{Fig1a}
    \end{subfigure}    
    \vspace{0.1cm} 
    \begin{subfigure}{0.7\textwidth}
        \centering
        \includegraphics[width=\textwidth]{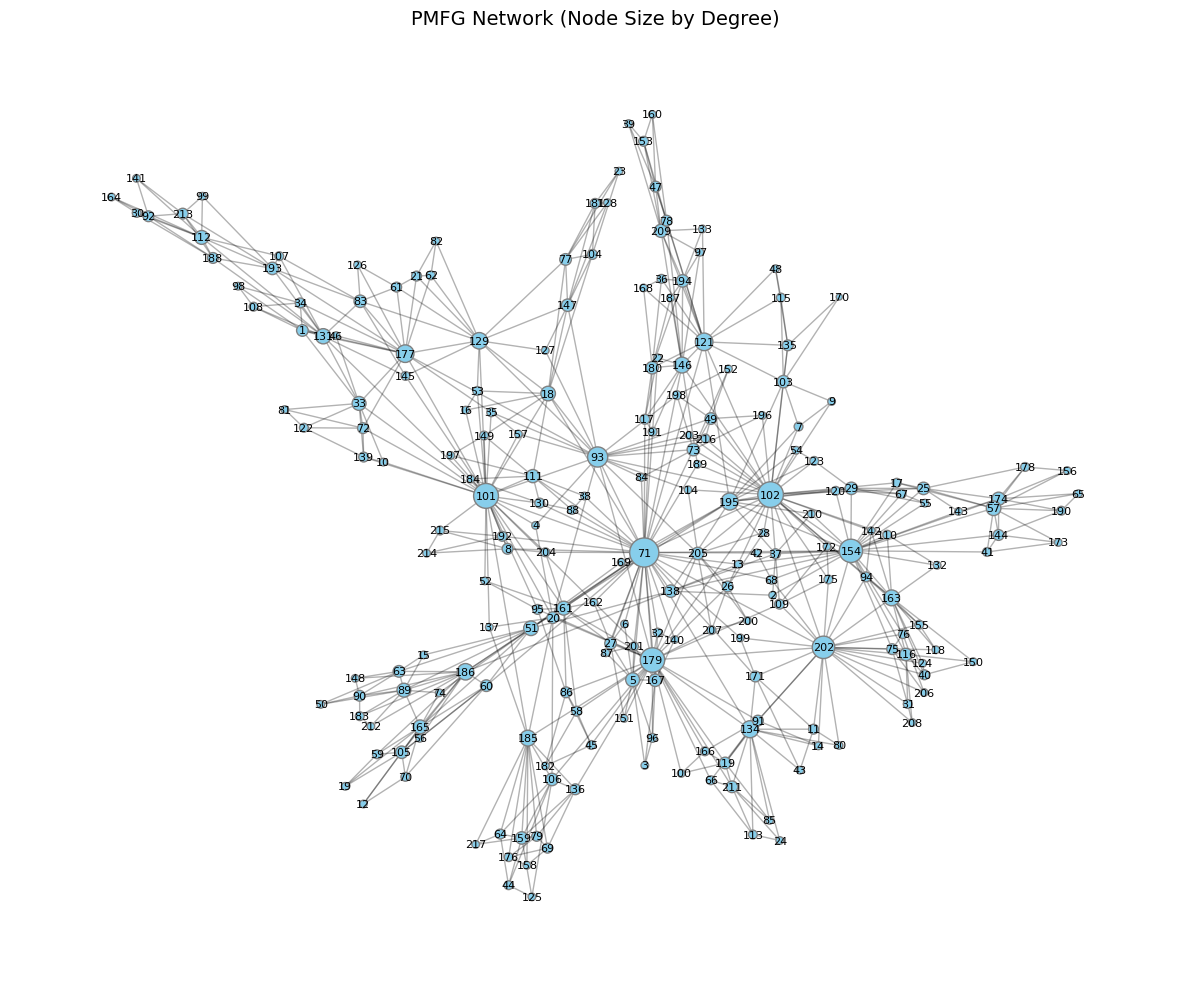}
        \textbf{(b)} 
        \label{Fig1b}
    \end{subfigure}
    
    \caption{(a) A network visualization for the year 2005, having an average network weighted degree of 4.10; (b) A network visualization for the year 2008, having an average network weighted degree of 4.55 (a). Note that both figures were generated using the weighted adjacency matrices, where node size represents the node's weighted degree.}
    \label{fig1}
    \vspace{2mm} 
    \textbf{Alt Text:} PMFG networks snapshots comparing the stock market in 2005 and 2008, where node size indicates degree. Both networks are based on weighted adjacency matrices, showing slight differences in average degree and diameter, with the 2008 network appearing slightly denser and more connected.

\end{figure}

\subsection{Complex Network Measures}
This section provides a concise overview of key classical network measures, focusing on edge-based measures such as edge betweenness centrality (EBC) and shortest path length (SPL). Then, we briefly state the mathematical definition of average clustering coefficient (ACC). We also introduce the concept of unweighted and weighted communicability measures and communicability distance. Within network science, a wide range of metrics derive from the notion of shortest path length, which is traditionally defined as the minimum number of edges required to connect two nodes, \(i\) and \(j\) \cite{boccaletti2006complex}. In the context of weighted networks, the shortest path length is computed as the minimum sum of edge weights across all possible paths connecting nodes \(i\) and \(j\), thereby accounting for each edge's varying strengths or costs.

\textbf{Edge betweenness:} The number of shortest paths that pass through an edge between two nodes is known as the edge betweenness \cite{newman2004finding}. This reflects the importance of an edge, facilitating communication between all nodes in the network. A node’s significance for the information flow throughout the network can be determined by applying the idea of betweenness to nodes as a measure of node centrality. Mathematically, the edge betweenness centrality \( C_{B}(e) \) for an edge \( e \) in a graph is given by:
\begin{equation}
C_{B}(e) = \sum_{i \neq j \in V} \frac{\sigma_{ij}(e)}{\sigma_{ij}}
\label{eq:5}
\end{equation}
where:
\begin{itemize}
    \item \( V \) is the set of vertices in the graph.
    \item \( \sigma_{ij} \) is the total number of shortest paths between vertices \( i \) and \( j \).
    \item \( \sigma_{ij}(e) \) is the number of shortest paths between \( i \) and \( j \) that pass through edge \( e \).
\end{itemize}
\textbf{Average Clustering Coefficient:} To quantify the tendency of nodes to form tightly knit groups within the network, we computed the average clustering coefficient (ACC), following the definition introduced by Newman (2003) \cite{newman2003structure}. This coefficient captures the ratio of the number of closed triplets (i.e., triangles) to the total number of connected triplets of nodes. Mathematically, it is defined as:
\begin{equation}
\text{ACC} = \frac{\sum\limits_{i,j,m} a_{ij} a_{jm} a_{mi}}{\sum\limits_i k_i (k_i - 1)}
\label{eq:6}
\end{equation}
Where $k_i$ is the degree of the $i^{th}$ Node in the network. This can also be computed for a weighted network, where edges represent the strength of the connection. 

\textbf{Communicability Measure:} The idea of communicability in a network is based on how a pair of nodes can communicate through walks connecting them. In usual settings, network measurements are based on the assumption that the information between two nodes travels over the shortest path, and communication between the two nodes is typically regarded as the shortest path. However, in many real-world networks, including financial, biological, social, and communication networks, information can travel down channels other than the shortest one and flow back and forth numerous times before arriving at its final destination. Indeed, especially in a network operating in a diffusion-like process, information does not always flow through the shortest paths because the sender may not know the network’s global structure: 
(i) it may not know which of the many routes connecting it to the addressee is the shortest, and (ii) even if it knows the shortest path, it may not know whether there are damaged edges along this path. In the literature, two definitions of communicability have been introduced: Estrada Communicability and Vibrational Communicability \cite{estrada2008communicability, estrada2012physics}. In this manuscript, we will restrict ourselves to Estrada Communicability \cite{estrada2008communicability}. Mathematically, it is defined between two nodes \(i\) and \(j\) of a network as: 
\begin{equation}
g_{ij} = \sum_{k=0}^\infty \frac{(A^k)_{ij}}{k!} = \left(e^A\right)_{ij}
\label{eq:7}
\end{equation}
Here, the $(i,j)$-entry of the $k^{th}$ power of the adjacency matrix \(A\) represents the number of walks of length \(k\) starting at node \(i\) and ending at node \(j\). The element \(g_{ij}\) encompassing all possible communication paths between two nodes, assigning greater significance to the shortest routes connecting them. It can also be understood as a measure of the probability that a particle, starting at node \(i\), will arrive at node \(j\) after randomly traversing the complex network. This communicability matrix is denoted as \(g\). Further, in the case of a weighted network, the communicability measure is adjusted by normalizing via node strength to avoid the heterogeneous connectivity standard in financial systems. This ensures that highly weighted nodes do not disproportionately dominate the communicability metric. The weighted communicability measure is defined as:
\begin{equation}
g_{ij}  = \left( \exp\left(S^{-1/2} W S^{-1/2}\right) \right)_{ij}
\label{eq:8}
\end{equation}
Where $S = diag(s_i)$ is the diagonal matrix, whose diagonal entries are the strengths of the nodes (i.e., $s_i  = \sum w_{ij}$). We will call this quantity-weighted communicability. 

The recent work by Diaz-Diaz and Estrada (2025) \cite{diaz2025signed} extends the communicability framework to signed graphs (adjacency matrix with both negative and positive edge weights). They demonstrate that the communicability matrix given by  Eq. (~\ref {eq:7}) remains valid for signed networks, with interpretation as: \(g_{ij}\) Encodes the net contribution of positive and negative walks, thereby capturing the effective synergy or interactions between nodes. In our work, we carried out the analysis based on both signed (Equation~\ref {eq:1}) and unsigned networks (adjacency matrix with non-negative edge weights). We constructed the unsigned networks by transforming the correlation matrix as $\tilde{C}_{ij} = \frac{1 + C_{ij}}{2}$, where \(C_{ij}\) denotes the original correlation from Eq. (~\ref {eq:1}). This is a standard approach widely adopted in the literature (see Lee et al., 2014 \cite{lee2014density}), as it preserves the structural properties and interpretability of the network while simplifying its representation. We observed consistent structural patterns and insights in both cases (signed and unsigned). For clarity and consistency, all analyses presented in this manuscript are based on the unsigned correlation networks.

\textbf{Shortest Communicability Path Length:}  The communicability distance \cite{estrada2012communicability} between a pair of nodes \(i\) and \(j\), denoted as \(\zeta(i,j)\), can be defined using the elements of Eq. (~\ref {eq:7}) as follows:
\begin{equation}
    \zeta(i,j) = \sqrt{g_{ii} + g_{jj} - 2g_{ij}}
    \label{eq:9}
\end{equation}
Where $g_{ii}$,  $g_{jj}$ are the self-communicability values, and $g_{ij}$ is the mutual communicability between nodes \(i\) and \(j\).  However, here we consider that “information” flows through the graph's edges in a network, using specific paths connecting the corresponding pair of nodes. Next, to find the shortest communicability paths between two nodes, we construct the communicability-weighted adjacency matrix of the network (Akbarzadeh \& Estrada, 2018) \cite{akbarzadeh2018communicability}:
\begin{equation}
    {X} = {\zeta} \ast \cdot {A} = {A} \ast \cdot {\zeta}
    \label{eq:10}
\end{equation}
Where \(A\) is the unweighted adjacency matrix of the graph, and \(.*\) indicates the element-wise operation. Next, the shortest communicability path length between two nodes is the shortest weighted path in the matrix \(X\). That is, the shortest communicability path between two nodes \(i\) and \(j\) is the one that minimizes the communicability distance across all possible paths connecting the nodes \(i\) and \(j\). Similarly, we can also calculate this communicability distance for the weighted communicability using Eq.(~\ref {eq:8}).
\subsection{Coalescent Embedding Algorithm}
This algorithm uses non-linear dimensionality reduction algorithms at its core. The dimensionality reduction is usually accomplished through popular algorithms like Isomap (ISO), Non-centered Isomap (ncISO), Laplacian eigenmaps (LE), Minimum curvilinear embedding (MCE), and Non-centered minimum curvilinear embedding (ncMCE) \cite{muscoloni2017machine}. All these algorithms embed the weighted network (with weighted adjacency matrix \(W\)) into the \(d\)-dimensional Euclidean space. However, these algorithms accept the pairwise distance between the nodes as input. Therefore, we convert the correlation-weighted adjacency matrix \(W\) into the corresponding correlation distance matrix \(M=[m_{ij}]_{n\times n}\) such that:
\[ m_{ij} = 
    \begin{cases}
        \sqrt{2(1 - C_{ij})}, & \text{if vertices } i \text{ and } j \text{ are adjacent} \\
        0, & \text{otherwise}
    \end{cases}
\]
Here, we choose the reduced low dimension of the embedded Euclidean space \(d = 2\). This step is called the dimensionality reduction (e.g., Isomap in our case) and approximates the geodesic distance between nodes. This distance matrix \(M\) is of order \(N\times N\). After the Euclidean embedding into the 2-dimensional space, we have, for each vertex \(i=1,2,3...N\), a pair of real numbers as a tuple \((x_i,y_i ) \in \mathbb R^2\).  Then, next, we assign the angular coordinates defined as:
\begin{equation}
    \theta_i' = tan^{-1}\left( \frac{y_i}{x_i} \right)
    \label{eq:11}
\end{equation}
Corresponding to each vertex \(i\). This approach to the angular assignment is referred to as the Circular Adjustment (CA). This is also called the original angular coordinates. The second method of angular assignment is the Equidistance Adjustment (EA). The \(i^{th}\) Vertex is defined by:
\begin{equation}
    \theta_i' = \frac{2\pi}{N} (t_i - 1)
    \label{eq:12}
\end{equation}
Where \(t_i\) are the ranks assigned to the \(i^{th}\) vertex, is derived by sorting the original angular coordinates in ascending order.

Lastly, the radial coordinates were assigned to each vertex in the network. First, the vertex degrees of the original network are arranged in descending order $d_1 \geq d_2  \geq ...\geq d_N$ (i.e., highest degree vertices appearing first). Then, the radial coordinate for each \(i = 1,2,...,N\) is chosen as:
\begin{equation}
    r_i = \frac{2}{\zeta} \left[ \beta \ln(i) + (1 - \beta) \ln(N) \right]
    \label{eq:13}
\end{equation}
Where \(\beta = \frac{1}{\gamma-1}\in (0,1]\) the power-law exponent. Thus, all the vertices of the network can, therefore, be represented by the polar coordinates \((r_i,\theta_i) \in \mathbb H^2\) in the Poincaré disc. Also, in light of \cite{dorogovtsev2000structure}, the above adjustment to the radial coordinate forces the vertices to the Poincaré disc. Thus, these polar coordinate transformation maps to the hyperbolic space and effectively interprets the popularity (radial) and similarity (angular) aspects of the original network in the Poincaré disc model. For a detailed study regarding the equation, refer to \cite{muscoloni2017machine, papadopoulos2012popularity}.

Further, we compute those above-defined measures over the embedded networks in the Poincaré disc model, by choosing the edge weight between vertices \(i\) and \(j\) as follows:
\begin{equation}
    w_{ij}= \frac{1}{1+x_{ij}}
    \label{eq:14}
\end{equation}
Where \(x_{ij}\) represents the hyperbolic distance metric between two nodes \(i\),  \(j\) in the Poincaré disc model. This hyperbolic distance for the two points having the polar coordinates \((r_i,\theta_i)\) and \((r_j,\theta_j)\) in the Poincaré disc is defined by 
\begin{equation}
    x_{ij} = \frac{1}{\zeta} \cosh^{-1} \left( 
        \cosh(\zeta r_i) \cosh(\zeta r_j) 
        - \sinh(\zeta r_i) \sinh(\zeta r_j) \cos(\Delta\theta)
    \right)
    \label{eq:15}
\end{equation}
Where, $\zeta = \sqrt{-K}$; We fixed $\zeta = 1$; $K$ the curvature of the hyperbolic space and $\Delta\theta = \pi - \left| \pi - \left| \theta_i - \theta_j \right| \right|$ is the angular distance between the nodes. Thus, we denoted these hyperbolic geometry-based measures as Hyperbolic edge betweenness (HEBC), Hyperbolic shortest path length (HSPL), and Hyperbolic communicability (HCOMM), and we will be using this notation to refer to them in the rest of the manuscript. This paper presents the results generated using the hyperbolic geometry-based measures for the Isomap-CA class of the coalescent embedding algorithm, since the results were similar for all other embedding classes.
\subsection{Sensitivity analysis on communicability}
The main objective of this study was to determine whether the communicability measure was a suitable metric to characterize the breakdown in communication between different pairs of stocks during the crisis or pandemic periods of the financial market. Therefore, we examine the sensitivity of the shortest communicability path length and shortest path lengths comparatively to detect significant changes in network connectivity during market crises. The procedure is outlined as follows:
\begin{enumerate}
\item Initially, for each connectivity matrix, we calculate the shortest communicability path length matrix (as directed in section 3.1) and the shortest path length matrix.
\item Next, we use the permutation test for the statistical significance analysis. Specifically, consider nodes \(x_i\)  and \(x_j\). Then, compute the shortest communicability path length between these nodes for each window corresponding to the periods of stability, denoted as \(l_{x_i,x_j}^{S_k}\), and each window corresponding to the periods of volatility, denoted as \(l_{x_i,x_j}^{V_k}\). Then, compute the respective average values of \(\bar{l}^S_{x_i,x_j}\) and \(\bar{l}^V_{x_i,x_j}\). After that, use these averages to calculate the difference:
\[\Delta l_{x_i,x_j}=\bar{l}^S_{x_i,x_j}-\bar{l}^V_{x_i,x_j}\]
\item To assess significance, randomization of windows was performed over all the combined windows of stable and volatile periods for that particular crisis. Generate 1,000 random subsets for stability period windows and 1,000 for volatility period windows. Next, for each random subset, compute:
\[\Delta l^{rand}_{x_i,x_j}=\bar{l}^{S_{rand}}_{x_i,x_j}-\bar{l}^{V_{rand}}_{x_i,x_j}\]
Where, \(\bar{l}^{S_{rand}}_{x_i,x_j}\) and \(\bar{l}^{V_{rand}}_{x_i,x_j}\) are calculated similarly but using the randomized groups.
\item Further, we compare the null distribution \(\Delta l^{rand}_{x_i,x_j}\) with the true value \(\Delta l_{x_i,x_j}\). A value \(\Delta l_{x_i,x_j}\) for a node pair \(x_i\) and \(x_j\) is considered significant if it lies outside the central 95\% of the null distribution. The \(p\)-value is calculated using permutation testing to confirm its significance.
    
\item A node pair is considered significant if its \(p\)-value is less than 0.001.

\end{enumerate}
 
\subsection{Classification Analysis}
The second objective of our investigation was to compare classification models trained using communicability versus models trained with standard metrics using the same dataset. Also, compare classification models trained using traditional network metrics versus models trained with network metrics obtained in the hyperbolic space. A linear support vector machine (SVM) \cite{apostolidis2015svm, james2013introduction} was used for learning and classification (For methodological details, see Appendix A). First, we start with calculating the weighted and unweighted adjacency matrices for both periods, as explained above. Then we compute the network measures of the original correlation-based network and the embedded network in the hyperbolic space. The edge weight between the nodes in hyperbolic space is settled using the approach (Eq.~\ref {eq:14}) referenced in \cite{cacciola2017coalescent,muscoloni2017machine}. The two types of network measures, topology-based measures (EBC, SPL, COMM) and geometry-based measures (HEBC, HSPL, HCOMM), were computed as explained in the above sections. They are referred to as ``connectivity measures'' in the remainder of the text.

Next, we perform the following stage, which involves correctly dividing the dataset into training and test sets to validate the classification process. This was accomplished by performing a 10-fold cross-validation 50 times. This procedure involved splitting the entire set of examples into ten subsets, or folds: one fold served as the test set, and the remaining folds comprised the training set. This splitting procedure was repeated until every fold was utilized as a test set once. Note that all combined windows were labelled by their corresponding stable group or volatile group. Additionally, we shuffled the windows into the folds for a more general approximation of the performance by repeating the 10-fold cross-validation 50 times with various permutations.

Furthermore, we performed the two-stage feature selection methods that were used to reduce the dimensionality of the feature space and ease the over-fitting issue. An ad hoc feature selection process applied to weighted connectivity matrices. To minimize the number of links into the network that needed to be considered for classification, corresponding to the stable period, a binary mean matrix, or a mask, was calculated, onto which the matrices of all windows (both periods) were subsequently projected (element-wise product of two matrices). The stable period mean matrix is a weighted matrix, where each entry $e_{ij}$   ranges from 0 to 1 and indicates the frequency of the corresponding edge that occurs among all the stable periods’ connectivity matrices.

We then threshold this mean matrix to obtain the referenced stable period binary mean matrix and use it as a mask matrix. The threshold value was chosen at 0.25, using a binomial test (with \(\alpha\) = 0.01): Assuming a prior probability of 0.5 that a link is present or not in a connectivity matrix, the binomial test established that a link is considered to be an ‘important’ link if it occurs in more than 25 \% of all stable period’s connectivity matrices. These stable period connectivity matrices show a more stable topology; volatile period connectivity matrices, instead, show a greater intra-variability due to the disrupted connectivity. This procedure would evaluate a significant and robust reference model to select the important links, as the sampling of windows considered in each round makes the definition of the set of significant links (i.e., the mask) robust with respect to outliers. Additionally, weak connections that can introduce noisy effects are filtered out in accordance with an objective statistical test with a strict significance threshold of the mask. This nested procedure of significant link selection was preferred to other threshold methods, such as fixing the same mean degree across all groups.

Using the feature matrices chosen in the first step, the second stage applied a traditional SVM recursive feature elimination (SVM-RFE) procedure \cite{guyon2002gene,lella2019communicability}, which is a powerful feature selection technique that utilizes the classification ability of the SVM to identify and retain the most relevant features for prediction. The criterion for feature selection used by SVM-RFE to rank the features based on the coefficients (weights) obtained from the trained linear SVM model. The feature with the least influence on categorization, as shown by its lowest ranking criterion, is eliminated at each iteration. Iteratively, this procedure keeps on until every feature is ranked. By keeping the best-ranked characteristics, feature selection is accomplished. This approach effectively reduces overfitting and enhances its generalization capability in high-dimensional spaces. For the detailed study regarding SVM-RFE, the reader may refer to \cite{guyon2002gene}.

To address underestimating strongly correlated features, we applied the method in \cite{yan2015feature}, which included a correlation bias reduction technique. To avoid feature selection bias and guarantee objective performance evaluation, a nested feature selection technique was used during cross-validation, in which the two-stage feature selection procedure was carried out only on the training set. An SVM model was trained to categorize data after dimensionality was decreased. The SVM algorithm uses its powerful generalization capabilities, which are especially useful in high-dimensional feature spaces, to find a separating hyperplane with the largest margin between classes. The summarized flow of this classification evaluation analysis is provided below in section 3.4.1.

\subsubsection{Algorithmic Steps for Classification Analysis}

\

\textbf{Step 1: PMFG Network construction and computation of connectivity measures}

\begin{enumerate}

    \item \textbf{Crisis Segmentation:} We consider two distinct financial crises for our analysis:
\begin{enumerate}[label=\Roman*.]
    \item \textit{Standard crisis:} Covering the years 2005 (stable) and 2008 (volatile).
    \item \textit{Unprecedented crisis:} Covering the years 2017 (stable) and 2020 (volatile).
\end{enumerate}
    
    \item We compute the Pearson correlation matrix (Eq. ~\ref {eq:1}) and construct the Planar Maximally Filtered Graph (PMFG) using a 60-day rolling window with 1-day sliding increments. This results in 132 PMFG networks for the stable Year 2005 and 124 for the volatile Year 2008. Similarly, we get 134 PMFG networks for the stable Year 2017 and 137 for the volatile Year 2020. Each PMFG network is also embedded into the hyperbolic space for geometry-based metric extraction.
    
    \item Compute the measure matrices (EBC, SPL, COMM) for each PMFG network and the geometric measure matrices (HEBC, HSPL, HCOMM) for the embedded PMFG networks of both crises’ periods.
    
    \item For each crisis separately, construct the mean matrix (with threshold 0.25) using the stable period as a mask matrix, onto which all the measure matrices of stable and volatile periods of that particular crisis are projected (an element-wise product of two matrices). We performed this separately for crisis I and crisis II.
    
    \item Feature vector extraction, each connectivity measure matrix is flattened into a feature vector of length $n(n-1)/2$, where $n$ is the number of stocks. Therefore, we have the complete data corresponding to two crises as follows: 
    \begin{itemize}
        \item Size of the standard crisis data: $256 \times 23721$
        \item Size of the unprecedented crisis data: $261 \times 73173$
    \end{itemize}
\end{enumerate}

\textbf{Step 2: Set up the SVM Model and Cross-Validation}

\begin{enumerate}
    \item Standardized dataset using z-score normalization via \texttt{StandardScaler} to ensure zero mean and unit variance for all features.
    \item Begin with the SVM classifier and use the linear kernel.
    \item Use repeated stratified k-fold cross-validation (5 splits, 10 repeats).
\end{enumerate}

\textbf{Step 3: Classification with Recursive Feature Elimination (RFE)}

\begin{enumerate}
    \item For each train-test split, split the standardized dataset into the train and test. Then, apply the RFE with SVM to retain the top feature.
    \item Then adjust both training and test sets to include only the selected features.
    \item Now fit the SVM using the new selected features of the training dataset.
    \item Now generate the predictions and the class probabilities.
\end{enumerate}

\textbf{Step 4: Performance Evaluation}

\begin{enumerate}
    \item Compute the standard metrics such as accuracy, AUC, sensitivity, and specificity. Then compare them by plotting the findings in a bar chart.
\end{enumerate}

\section{Results}
The findings of the statistical analysis on communicability and SVM classification are described in the following subsections.
\subsection{Communicability and Structural Dynamics of Financial Networks During Distinct Crises}
\begin{figure}[H]
    \centering
    \begin{subfigure}{0.999\textwidth}
        \centering
        \includegraphics[width=\textwidth]{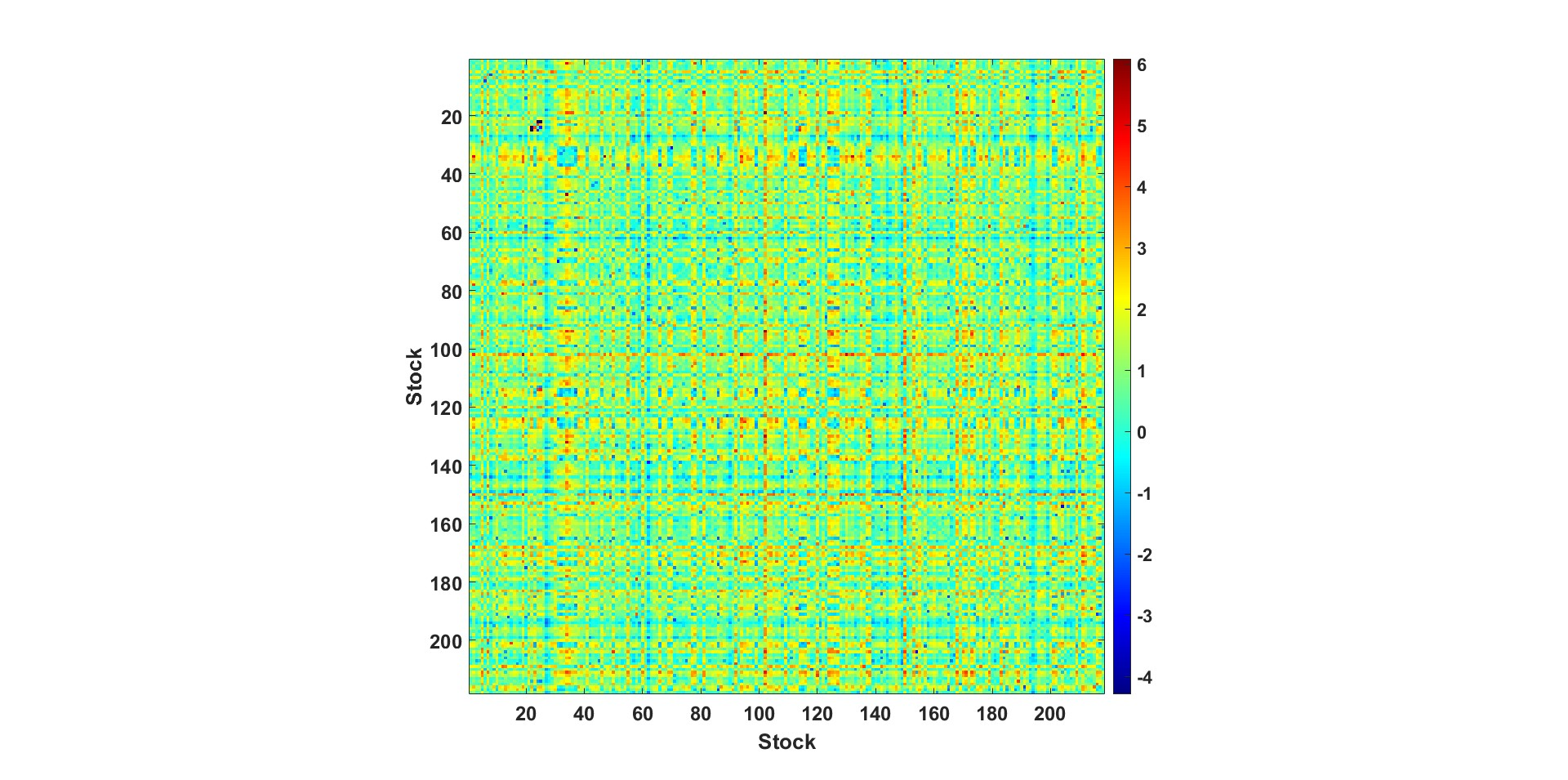}
        \textbf{(a)} 
        \label{Fig2a}
    \end{subfigure}    
    \vspace{0.1cm} 
    \begin{subfigure}{0.999\textwidth}
        \centering
        \includegraphics[width=\textwidth]{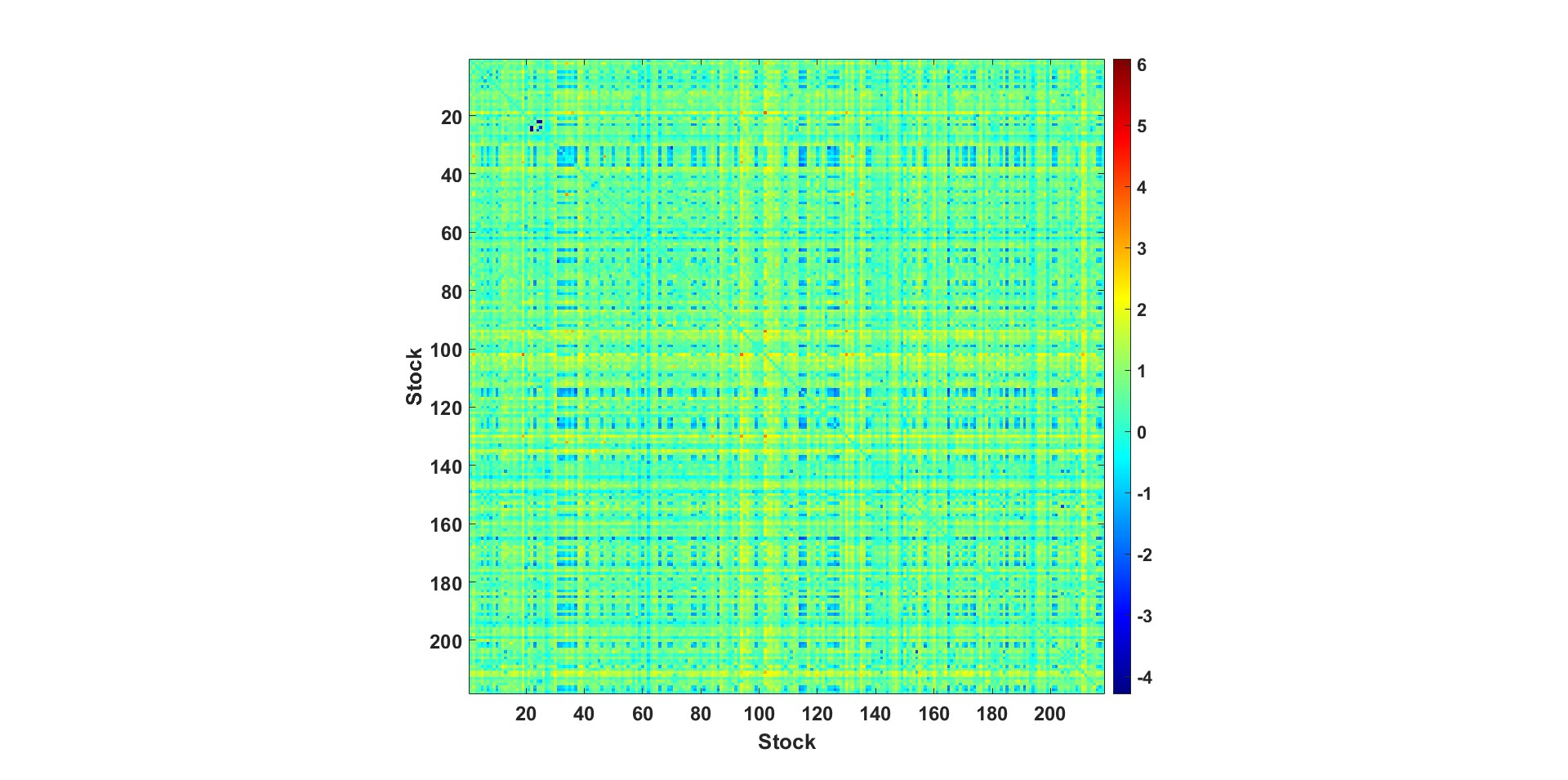}
        \textbf{(b)} 
        \label{Fig2b}
    \end{subfigure}
    
    \caption{(a) Heat map of the difference between the averaged shortest communicability path length for the periods of stability and volatility; (b) Heat map of the difference between the averaged shortest path length for the periods of stability and volatility, on the same scale as (a). Both figures are generated using the unweighted adjacency matrices corresponding to both periods.}
    \label{Fig2}
   \vspace{2mm} 
   \textbf{Alt Text:} Heatmaps showing the differences in average shortest communicability path length and shortest path length between stable and volatile periods, respectively, highlighting the sensitivity of communicability measures to structural changes in network connectivity.
\end{figure}

To uncover the structural patterns in financial networks during major crises, we worked with two distinct time windows for each crisis event: a stable and a volatile period. For the global crisis of the year 2008, we considered the stable (2005) and volatile (2008) periods and computed the average shortest communicability path length matrix for both the periods of stability and volatility. Next, we have derived the difference between the matrix corresponding to the stability and volatility periods provided as a heat map in Figure ~\ref{Fig2} (a). We also calculated these differences for the shortest path lengths between each pair of nodes for comparison’s sake, Figure ~\ref{Fig2} (b). The differences for the shortest path length were set at the same scale as the differences of the shortest communicability path length, and we represented these differences using a diverging colour map centered at zero. These figures show that the communicability path length is more sensitive to the connectivity in the stock pairs than the shortest path length. Similarly, we plot those two figures for the pandemic crisis period of 2020. The heatmap figure of the difference between the average shortest communicability path length matrix and the average shortest path length matrix corresponding to the stability (2017) and volatility (2020) periods is shown in Appendix B (see Fig. B4), capturing the market sensitivity in the unprecedented crisis.

Notably, we observed in the communicability difference heat maps that the average shortest communicability path length of most of the banking stocks, such as AXISBANK, PNB, ICICIBANK, SBIN, and BOBANK, is positive with most of the other stocks than the communicability between the other stock pairs. This suggests that these stocks exhibit a higher level of connectivity with other stocks during the crisis, potentially reflecting their central role in the correlated market structure under volatile conditions. Thus, it is also consistent with the existing literature on the stock market.

\begin{figure}[H]
    \centering
    \begin{subfigure}{0.99\textwidth}
        \centering
        \includegraphics[width=\textwidth]{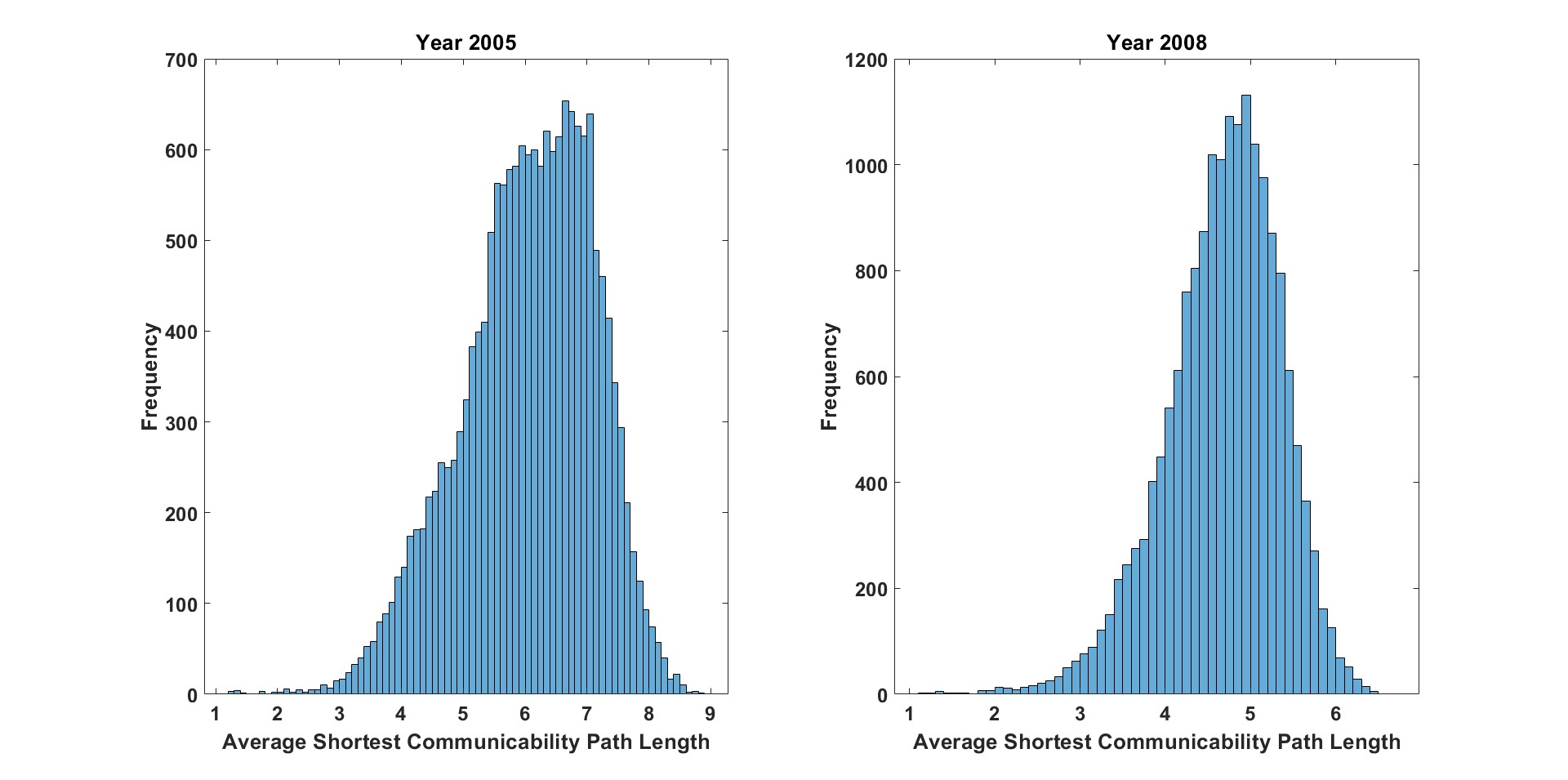}
        \textbf{(a)} 
        \label{Fig3a}
    \end{subfigure}    
    \vspace{0.1cm} 
    \begin{subfigure}{0.99\textwidth}
        \centering
        \includegraphics[width=\textwidth]{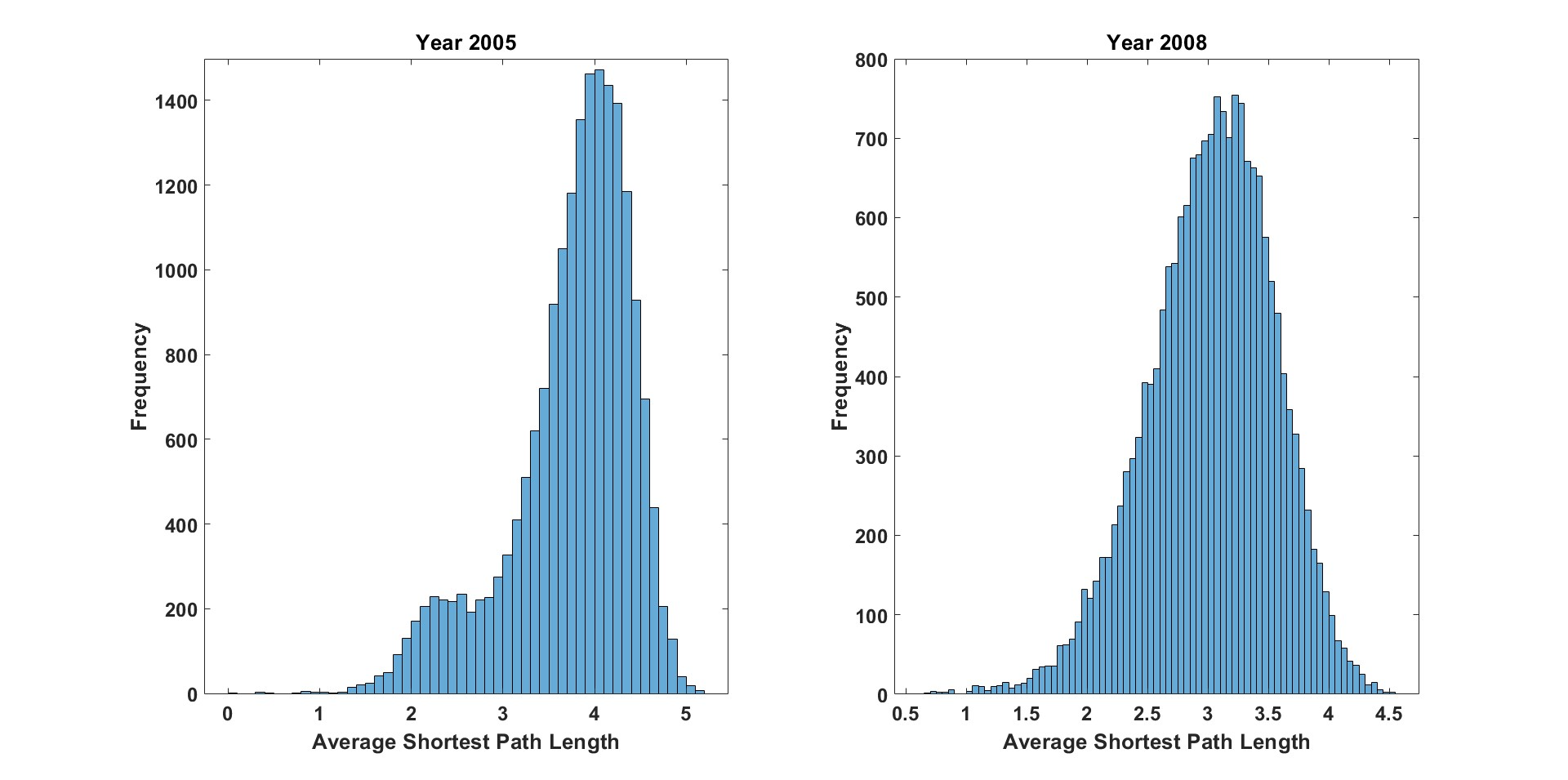}
        \textbf{(b)} 
        \label{Fig3b}
    \end{subfigure}
    
    \caption{(a) Histogram of the average shortest communicability path length in the stable period (Year 2005) and the volatile period (Year 2008) corresponding to the significant stock pairs, respectively (b) Histogram of the average shortest path length in the stable period (Year 2005) and the volatile period (Year 2008) corresponding to the significant stock pairs, respectively. Both figures are generated using the weighted adjacency matrices corresponding to both periods.}
    \label{Fig3}
    \vspace{2mm} 
    \textbf{Alt Text:} Histograms comparing average shortest communicability and shortest path lengths for significant stock pairs during stable (2005) and volatile (2008) periods, highlighting a negative shift in both the measures from stable to volatile indicates a clear structural change in significant stock pairs.
\end{figure}

Additionally, we performed the permutation test procedure outlined in section 3.4 on the stable and volatile periods. For the global crisis of the year 2008, we found 73.3\% of the total distinct stock pairs have statistically significantly different communicability (adjusted $p<$ 0.001). Notably, most of these significant stock pairs showed a decrease in average shortest communicability path length in the volatile period. Further, looking at the distribution of significant stock pairs of the average shortest communicability path length values in Figure ~\ref{Fig3} (a), a hypothesis of a decrease in population median in the volatile period was tested by performing a one-sided Wilcoxon rank sum test \cite{zar1999biostatistical}. A \(p\)-value of 0 was obtained: with enough evidence, it can be concluded that there is a negative shift in the median of the average shortest communicability path length values of the significant stock pairs at the 0.001 significance level. This pronounced shift in the average shortest communicability path length from 2005 to 2008 indicates a clear structural change in significant stock pairs. In 2005, values were centred around a higher mean of 6.05, implying less communicative and loosely connected, while in 2008, values were centred more around a lower mean of 4.76, indicating a stronger overall communicability and denser connectivity, likely due to heightened market synchronization during the 2008 crisis. This shift in central tendency is more evident in the average shortest communicability path length than in the average shortest path length.

A similar analysis was performed over the pandemic crisis of 2020. We found that around 80\% of the total distinct pairs were found to have statistically significantly different communicability (adjusted $p<$ 0.001). However, in contrast to the global crisis of 2008, during this pandemic, most of these significant stock pairs showed an increase in the average shortest communicability path length in the volatile period of 2020. This shift was further claimed by looking at the average distribution of the shortest communicability path lengths during stable and volatile periods, as shown in Appendix B (Fig. B5). A one-sided Wilcoxon rank sum test tested the hypothesis of an increase in population median in the volatile period. This shows a positive shift in the population median during the volatile periods with a 0.001 significance level. This demonstrates that overall communicability and connectivity disruption occurred during the pandemic crisis, which was the opposite of the global crisis of 2008. These contrasting patterns highlight fundamental differences in the nature and impact of the two crises. While the 2008 financial crisis led to systemic communicability and connectivity amplification within the market network, the 2020 pandemic, in contrast, was associated with a reduction in the global communicability in the significant stocks; this implies that indirect links and overall ease of information flow across the network became significantly less efficient in 2020. Such a shift is consistent with the market being under stress, possibly caused by the COVID-19 pandemic, which led to heightened volatility, uncertainty, and unprecedented market reactions. This highlights the context-dependent nature of financial markets during crises and underscores the value of communicability metrics in capturing such structural information in times of volatility.
\subsection{Classification Performance Evaluation}
\begin{figure}[H]
    \centering
    \includegraphics[width=0.8\textwidth]{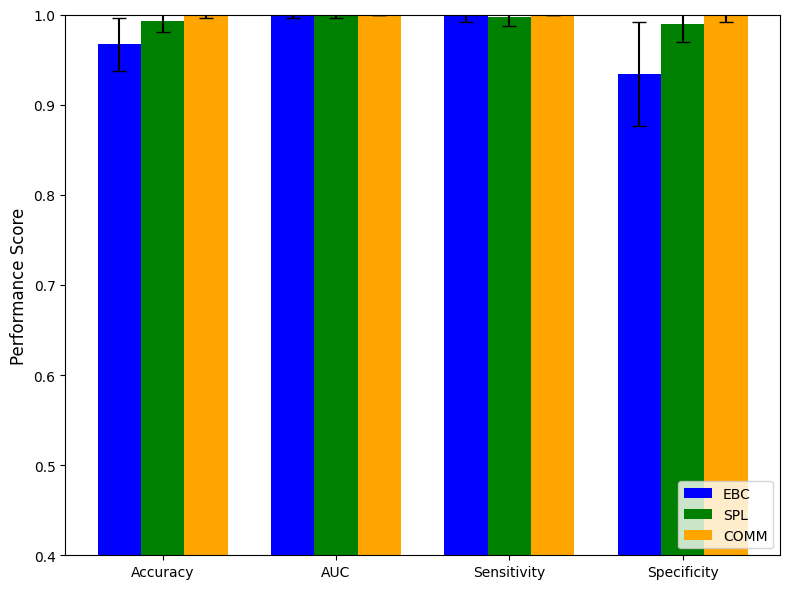}
    \caption{Overview of the performances achieved using the individual topology metrics computed on the weighted connectivity matrices as features for the stable and volatile periods classification.}
    \label{Fig4}
    \vspace{2mm} 
    \textbf{Alt Text:} Comparison of classification performance metrics for topological measures applied to weighted stock market networks, distinguishing between stable and volatile periods of the crisis 2008.
\end{figure}

\begin{figure}[H]
    \centering
    \includegraphics[width=0.8\textwidth]{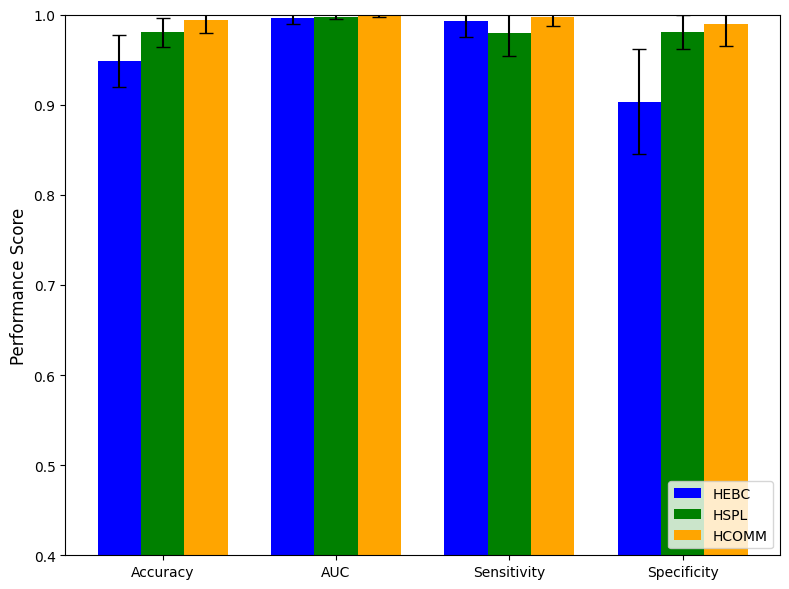}
    \caption{Performance overview of all the connectivity measures versus the hyperbolic connectivity measures, computed on weighted connectivity matrices (stable and volatile periods classification).}
    \label{Fig5}
    \vspace{2mm} 
    \textbf{Alt Text:} Comparison of classification performance metrics for geometric measures applied to weighted stock market networks, distinguishing between stable and volatile periods of the crisis 2008.
\end{figure}

We used the four natural performance measures, namely accuracy, sensitivity, specificity, and area under the ROC curve (AUC), to report the model classification performances (For the details regarding Evaluation measures, see Appendix A), which are averaged throughout all cross-validation rounds. First, we performed the SVM model for the classification between the periods of stability and volatility of the global crisis of the year 2008. Figure ~\ref{Fig4} summarises the classification result attained with each topology-based connectivity measure separately (SPL, EBC, COMM). Each performance metric represents the mean value derived from averaging the outcomes across all cross-validation iterations. Each bar’s height in the figure represents the average performance, while the error bars show the corresponding standard errors. Overall, each topology measure performs superbly in classifying stable and volatile periods. Each topology measure scores more than 0.94 in all four performance measures. Notably, the COMM measures perform accurately in all four performance measures for the classification between the periods and outperform the shortest path-based measures, SPL and EBC. We also observe that SPL is slightly better than EBC in classification. 
Figure ~\ref{Fig5} shows the results corresponding to geometric measures such as HEBC, HSPL, and HCOMM for classification between the periods of stability and volatility. These geometric measures also perform well in classifying between the periods. We also observe that geometry-based measures perform comparably to the topology-based measures in classification during the 2008 crisis.
A similar experiment was conducted for the 2020 pandemic crisis to classify periods of market stability (2017) and volatility (2020). The relevant results for both topology-based and geometry-based measures are presented in Appendix B (see Fig. B6 and Fig. B7). Both sets of measures demonstrated strong classification performance, effectively distinguishing between stable and volatile periods. However, the geometry-based measures exhibited a slight decline in performance compared to the topology-based measures. This may be attributed to their higher effectiveness in capturing complex, nonlinear structures, which may not be as prominent in this dataset. Overall, communicability measures outperformed edge-based shortest path metrics in identifying periods of stability and volatility across both crises, which are different in nature.

We performed a surrogate data analysis to ensure that our classification methodology and results are not artifacts of circular reasoning or driven solely by marginal statistics such as volatility. The dataset was created by independently shuffling the return time series of each stock within every time window, thereby preserving individual stock volatility while destroying the cross-stock correlations that underpin the network structure. This is a standard approach explored in the work (Araujo et al. 2007) \cite{araujo2007geometry} to test whether the observed patterns in the financial time series reflect genuine structure or are merely the result of random fluctuations. We applied this procedure to the 2008 global economic crisis and the 2020 market shock period. We observed a substantial drop in classification performance, approaching a baseline score of 0.5 under the surrogate condition. This decline in the model performance confirms that the original classification is based on the meaningful topological features captured by the network metrics, rather than noise or simple statistical properties. The results of this analysis provide strong evidence that our predictive metric features genuinely capture structural shifts in the financial system.
\section{Discussion}
The aim of this study was to examine the financial correlation-based network and demonstrate how the communicability measure may be used to highlight changes in connectivity between stock pairs during a stock market crisis. The advantage of using this metric has been evaluated from two points of view:
\begin{itemize}
    \item A statistical analysis revealed that stock pairs have significantly different communicability values during volatile and stable periods.
    \item	The first observation in the classification methodology demonstrates that the shortest path-based network measurements can almost precisely differentiate between market stability and volatility periods.
\item	Secondly, the classification framework with binary groups of features showed how communicability positively affects classification performance, both for the stable period/volatile period.
\item	The hyperbolic distance-based metrics perform comparably to the original correlation-based metrics regarding classification robustness and accuracy. Furthermore, these two kinds of measures demonstrate that the communicability has a greater favourable impact on classification performance than the shortest path-based measures.
\end{itemize}
First, 70.3\% and 80\% of stock pairs had statistically significantly different communicability values during the systemic crisis of 2008 and the unprecedented crises of 2020, respectively. Second, in the classification scenarios, this study demonstrates that when features are incorporated in the classification model, the shortest path-based network measures and the communicability measure almost reliably distinguish between market stability and volatility times. Furthermore, it has been demonstrated that in the financial system, communicability appears to be a more effective measure than those based on shortest paths. Finally, it was shown that the geometry-based measures were comparable to classifying the market periods with topology-based measures. According to the existing literature, changes in network connectivity measures between stable and volatile periods are expected to occur. However, we notice that the communicability provides us with distinct advantages. Unlike the measures based solely on shortest path distances, communicability includes all weighted walks between nodes, providing a richer structure of systemic connectedness. Its heightened sensitivity during crisis periods, as evidenced by our statistical and classification analyses, suggests that communicability may serve as a more effective metric for early detection of market-wide stress and systemic risk. 

Interestingly, our findings also show the contrasting communicability behaviour across the distinct crises: a decrease in the 2008 global financial crisis and an increase in the pandemic turmoil of 2020. This divergence may indicate the different underlying mechanisms of those crises. The 2008 crisis lies in the structural failure of the system,  likely due to disruption in the connectivity of information flow among the stocks, while on the other hand, the COVID-19 crisis was caused by the unprecedented health shock, which led to an increase in market-wide uncertainty, more synchronized co-movement, increased interdependency reflected in the shift in the communicability measures. This contrast behaviour highlights and explains how the nature and origin of the crisis fundamentally affect the dynamics of the financial networks.
\section{Conclusion}
In this study, we explored the role of the shortest path-based network measures and the communicability measures in characterizing structural changes in the financial networks, specifically across market stability and volatility periods. These measures successfully discovered the connectivity differences between these periods. The communicability measure provides a nuanced understanding of the financial network and can be considered an alternative measure to traditional shortest path-based measures.

The primary goal of this article was to introduce a novel methodology in the classification analysis for studying the stock network during different market conditions. We begin by highlighting the importance of investigating metrics that better capture the diffusive processes in financial systems. To this end, this study initially utilized the communicability with a statistical-descriptive objective: stock pairs with statistically significant differences in communicability values during volatile and stable periods were identified. This indicates the enhanced connectivity and the heightened information flow between the stock pairs during the turbulence periods. In the second step, the advantage of applying communicability was investigated from a quantitative point of view: it was demonstrated that the classification tasks between the periods of stability and volatility, individually and along with the communicability measure as features for training the classification models. The study reports that the network measures can individually classify the market periods almost accurately, while the communicability measure outperforms the traditional network measures. Next, this study shows that geometry-based measures have a classification capability comparable to network topology-based measures. 

The efficiency of the communicability measures in uncovering connectivity differences between the two different stock market periods convinces us that communicability could be a decisive distinguishing factor for a more accurate understanding of the market information dynamics. Using the communicability measure in this context can represent a starting point for developing new classification strategies based on this measure that could further improve classification performance. The advantage of using communicability for this purpose is not limited to classifying the financial periods. It can also contribute to the early detection of the stock market crisis, which is one of the significant challenges in current research in the finance system. In summary, this study combines machine learning, network science, hyperbolic geometry, and financial analysis to provide fresh and insightful perspectives on the crises in the Indian stock market. It also presents an approach that successfully predicts and classifies the periods of market volatility from stability. The proposed approach demonstrates high classification performance and emphasizes the comparable performance of the geometry-based network measures. 
Although our findings contribute to a deeper understanding of the stock market structure and reveal several interesting insights. While it also has certain limitations. Since it relies on empirical data from the NIFTY 500 Index of the NSE, which may not fully represent the broader Indian or global financial markets, it potentially limits the generalizability of the findings. Next, this analysis focuses on the Planar Maximally Filtered Graph (PMFG) network, excluding other network models like the PMFG-based threshold network (PTN), the threshold-based network, which could offer complementary insights. This study also designates stable and volatile periods based on standard deviation. Still, this binary classification may oversimplify market dynamics, overlooking intermediate states or other volatility drivers. 

In the future, we can extend this methodology and conduct it in other foreign stock markets, such as the Shanghai and New York stock exchanges, to observe the consistency and robustness of this work outside of the Indian stock market. An important future direction is to implement the generalized shortest path framework proposed by Opsahl et al. (2010) \cite{OPSAHL2010245}, which utilizes both the number of connections and the strength of those connections in computing shortest paths. This dual-weighted method addresses the limitation of assuming that smaller weights always represent stronger or more efficient connections. The future direction of this work will explore how this approach may alter the identification of central nodes or paths in financial crisis periods. Additionally, it would be interesting to explore and relate the firm-level characteristics, such as sectoral information, valuation, and market capitalization of stocks, with the communicability dynamics. This kind of analysis may guide us in determining which firm derives the systemic inter-dependencies, potentially enhancing the application in tracking the contagion and sector-level risk. Thus, this work opens up several potential directions that could provide a deeper understanding of the financial market’s dynamics and structural characteristics.

\section*{Acknowledgment}

We thank the Shiv Nadar Institution of Eminence for providing the computational resources and infrastructure to conduct this research. We thank Professor Sanjeev Agrawal and Dr. Qazi Azhad Jamal for their support and helpful comments. The first author thanks the Council of Scientific and Industrial Research (CSIR), India, Senior Research Fellowship Scheme (Grant No. 09/1128(12964)/2021-EMR-I) for financial assistance for this study.


\bibliographystyle{comnet}
\bibliography{sample}
%








\end{document}